# Towards completing Planetary Systems:
# The role of minor bodies on life growth and survival

Jorge Lillo-Box (European Southern Observatory - ESO, jlillobox@eso.org), David Kipping (Columbia University), Isabel Rebollido (Universidad Autónoma de Madrid, UAM), Pedro Figueira (ESO), Adrien Leleu (Bern University and Observatoire de Paris), Alexandre Correia (Aveiro University), Philippe Robutel (Observatoire de Paris), Nuno C. Santos (IA, Porto), David Barrado (Astrobiology Center, INTA-CSIC), Benjamín Montesinos (Astrobiology Center, CSIC-INTA), Tjarda Boekholt (Aveiro University)

**Abstract**

The search for extrasolar planets in the past decades has shown that planets abound in the Solar neighborhood. While we are still missing an Earth twin, the forthcoming space missions and ground-based instrumentation are already driven to achieve this goal. But, in order to fully understand the conditions for life appearing in the Solar System, we still miss some pieces of the planetary system jigsaw puzzle, namely a deeper understanding of the minor bodies. Trojans, moons, and comets are tracers of the formation and evolution processes of planetary systems. These missing pieces are also critical to understand the emergence and evolution of life over millions of years. With the large crop of planetary systems discovered so far and yet to be detected with the forthcoming missions, the hunt for minor bodies in extrasolar systems is a natural continuation of our search for real Solar System- and, in particular, Earth- analogs. This white paper is focused on detection of these minor components and their relevance in the emergence, evolution and survival of life.

## Life in planetary systems: the role of minor bodies

Planet Earth is currently the only one we know to sustain life on its surface. Since the detection of the first extrasolar planets more than 20 years ago, the technological progress and the development of new techniques have not only boosted the number of known extrasolar worlds but is also approaching the detection of Earth analogs. In our search for life beyond our planet, looking for these rocky worlds in the habitable zone of their host stars is only the first step of a long interdisciplinary ladder. The few thousands of extrasolar planets found so far have shown the broad variety of physical properties and architectures that planetary systems can exhibit. This plethora is a direct consequence of the many factors involved in planet formation and early evolution (i.e., migration, great impacts, etc.) which can mold the shape of the systems in many different (and somehow chaotic) ways.

We know that kilometer-size bodies are direct outcomes of planet formation regardless of the formation mechanism itself (e.g., Safronov & Ruskol, 1992). These objects influence the environment in many different ways, including transportation of material (e.g., water) from different parts of the system and potentially depositing them in the surface of planets by direct collisions in the case of asteroids/comets, polluting the stellar surface, influencing the surface components of the planets in the case of moons, or potentially reshuffling the masses of protoplanets in great impacts in the case of trojans. In our own Solar System, we have several examples about how these minor components played (or can play) a key role in the growth, evolution and survival of life on Earth as well as potentially hosting life themselves. Here we summarize some examples in the Solar System:

- Life on moons: The possibility of Solar System moons hosting life on their surface (Titan) or in big oceans below the frozen crust (Enceladus) is currently being analyzed from different perspectives (e.g., Palmer et al., 2017).

- The effect of the Moon on the Earth: On Earth, the obliquity variations due to planetary perturbations are confined within 22 - 24 degrees, but yet they resulted in impressive climatic changes during the past, known by the Milankovitch theory of paleoclimate (Milankovitch 1941). However, these small obliquity variations are only possible because of the presence of the Moon (Laskar, Joutel & Robutel, 1993). Indeed, the gravitational pulling of the Moon increases the precession rate of the Earth, which prevents resonant interactions with the other planets. In the absence of the Moon, the obliquity of the Earth would undergo chaotic variations from 0 no 60 degrees like those presently observed for Mars (Laskar & Robutel 1993). These changes would result in catastrophic global changes in the Earth's climate, that would likely prevent that life emerges as we know it.

- Trojans shaping the architecture of planetary systems: A Mars-sized trojan is one of the potential origins of the impactor that formed the Moon in the Great Impact theory (e.g., Hartmann & Davis, 1975; Cameron & Ward, 1976; Belbruno & Gott, 2005).

- Rossetta: the transportation of water in the Solar System from the outer regions to the rocky regime is possibly driven by comets and asteroids traveling along the whole planetary system and depositing their contents during impacts with the rocky planets.

As shown in these examples in our Solar System, the detection of minor bodies in extrasolar planetary systems (together with their in situ study in the Solar System) is critical to have the complete picture of planetary systems and their possibilities to host life. Some examples are:

- The physical, chemical and dynamical properties of trojan bodies and even their mere presence or absence depend on their formation mechanism and can thus be a proof of planet migration processes (e.g., Beaugé et al. 2007; Cresswell & Nelson 2009). Also, trojans in stable orbits corevolving with gas giants in the habitable zone of their parent star are potential targets to support the conditions to grow life (Dvorak et al., 2004).

- Comets, as well as asteroids, are the leftovers of planetary formation processes, and so they carry fundamental dynamical and chemical information about the evolution of the system. The study of comet composition may shed light on the origin of water and molecules on Earth (e.g., hypothesis of S. Arrhenius; Wickramasinghe, 2015 and references therein).

- Understanding $\eta$-Earth requires measuring $\eta$-Moon. It is not enough to put an Earth-size planet in the habitable zone to ensure a proper environment for life. The obliquity of a planet (the angle between the equator and the orbital plane) determines the meridional insolation gradient at surface. By increasing the obliquity we reduce the insolation on the equator and increase it at the poles and vice-versa.

## Progress since "New Worlds, New Horizons"

The hunt for exomoons started more intensively around 10 years ago and was boosted by the launch of the Kepler mission (e..g, Kipping 2009; Kipping et al., 2012; Heller et al., 2014). Constraints on the presence of Gailean-sized moons interior to 1 AU is one of the results of this search, together with the finding of the best exomoon candidate found so far, Kepler-1625 b-i (Teachey et al. 2018). When it comes to trojans, the large amount of data gathered so far to detect extrasolar planets and to precisely measure their properties has also allowed a search for trojan bodies using different techniques (e.g., Ford et al., 2006, 2007; Leleu et al., 2015, 2017). But, the different attempts done so far have not been fruitful in finding co-orbitls up to a few tens of Earth masses (e.g., Ford et al. 2006; Madhusudhan et al., 2008; Leleu et al., 2017; Lillo-Box et al., 2018). Unlike exomoons and exotrojans, exocomets have been detected for the last 30 years since its first discovery by Ferlet et al. (1987) in Beta Pictoris. Ever since, metallic photospheric (mainly CaII or NaI) lines have been analyzed in the search for superimposed narrow variable absorptions, correspondent to the gaseous tail of comets developed when the frozen bodies approach the star (Beust et al., 1990). Around 20 systems showing variable absorptions that can be attributed to exocometary activity have been reported (Eiroa 2016, Welsh & Montgomery 2017, and references therein).

The presence of hot and cold gas in the system, as well as evidence of material surrounding the star such as the infrared excesses produced by debris disc can also probe the presence of exocometary material (Rebollido 2018).

**Key scientific questions for the next decades**

Despite the progress done so far and very briefly summarized in the previous section, still many scientific questions remain unanswered. In most cases, this is due to the inability of current technology to easily detect these minor bodies as we have started to do with planets. In the following, we identify some of the most relevant scientific challenges related to the three main types of minor bodies present on planetary systems that remain unsolved and that would need intense observational and theoretical effort within the next decades.

**Exomoons:**

- Formation pathways for large moons.- Regular satellites, such as those of Jupiter, are thought to have formed in-situ, which is hypothesized to limit their growth to ~10,000$^{th}$ the mass of the planet (Canup & Ward 2002), but an extrasolar observational test of this remains absent. Other mechanisms clearly exist too, such as captures and impacts (Agnor et al. 2010). Here, it is not well-understood how large such moons could become. Extrasolar examples would strongly inform the viable pathways building large, potentially habitable, moons and even the possibility of binary planets.

- Survival statistics.- It has become clear that both within the Solar System and beyond, planetary migration is likely a universal and critical process behind the observed architectures. The survival of moons in such systems is far less well studied and empirical studies using Kepler data suggest that Gailean-sized moons seem not to be common interior to 1 AU (Teachey et al. 2018). Under what conditions and via what mechanism are they lost? What possibly multi-variate boundary which delineates their survival?

- Exomoon properties.- In the Solar System, the majority of moons have higher ice contents than that of the terrestrial planets due to their location beyond the snow line. Further, only Titan possesses a thick atmosphere. Are these patterns typical? Almost everything about the planets in the Solar System appear to buck the trends of typical exoplanets and there appears no clear basis to believe exomoons will not similarly surprise us. Also, tidal heating plays a central role for moons such as Io and Europa and may even permit habitability under certain conditions. Could tidally heated moons be another unusual quirk of the Solar System?

**Exotrojans:**

- Formation.- Several formation mechanisms have been proposed in the past decades.(mainly in situ formation and capture; see Laughlin & Chambers, 2002, and references therein). However, none of them seem to be predominant in the Solar System. So, what is the predominant mechanism (if any) in general? When are they formed?

- Properties.- The properties of trojan bodies are presumably related to their formation mechanism. Is it possible to form planet- or moon-size trojans? Under which conditions can they remain stable? Can we expect larger trojans in other systems?

- Habitability.- Trojan co-orbiting with massive planets in the habitable zone (currently some tens of planets discovered are in the conservative and optimistic HZ) could potentially be also habitable (e.g., Dvorak et al. 2004; Schwarz et al., 2016). Several additional questions need to be addressed before considering trojans as potential habitable worlds. Assuming they can grow up to moon or planet-size, can they form and retain an atmosphere? what would be the source of the atmospheric components (e.g., impacts with other smaller trojans, outgassing from vulcanism processes, etc.)? At first, these questions need to be addressed theoretically. Observationally, the detection of the first exotrojans and their atmospheric characterization (as we will do with Earth-analogs using the forthcoming instrumentation as JWST or E-ELT) can provide hints to theoretical models. Additionally, trojans can be trapped into spin-orbit resonances different from the synchronous state thanks to the co-orbital motion. This implies that even if the habitable zone of a star is located within the area when we believe that the planet would be tidally locked in a 1:1 spin-orbit resonance (like most of TRAPPIST-1 planets), the co-orbital motion can actually allow the repartition of the insolation on the entire surface of the planet (creating day/night cycles), see Correia & Robutel (2013) and Leleu et al. (2016).

**Exocomets:**
- Formation and evolution: Comets might be a result of either the interruption of pebble accretion process or the leftovers of a destructive collision between planets or protoplanets. But, when do comets appear the in the systems and how long they survive?
- Exocometary environments: Exocomets are found around 20 stars, most of them A-type. Is this an observational bias? Further investigation in a range of stellar temperatures and also theoretical models of comet survival and gas removal are still needed. Also, the necessity of photometric follow up of those targets could be satisfied with the new upcoming missions for exocomet survey. A combined spectroscopic and photometric observations of exocomet-bearing stars in order to better constrain the system's properties would be extremely helpful since transits allow a better determination of the mass of the comets, while spectroscopic follow up could determine the orbits and the composition.
- Upcoming missions : Since exocomets are only found in bright, hot stars, they can't be observed with Kepler or Gaia. Upcoming missions like TESS and Plato, will allow a much lower magnitude range (up to 4 mag), making it possible to observe some of these stars in photometry from space. Also, JWST will allow observations of much fainter debris disc, since its much larger aperture increases several orders of magnitude its sensitivity. Ground based telescopes like E-ELT harboring high-resolution spectrographs will also enhance the sample of exocomet-bearing stars. These new missions will allow to investigate a possible relation between the presence of exocomets and planets.

## Key technological challenges for the next decades

In order to achieve the above scientific goals, besides the theoretical works, new instrumentation and new technology is required to routinely detect these minor bodies in extrasolar planetary systems. Given the expected sizes of exomoons and exotrojans (from Moon to Earth radii), we foresee the transit technique to be the most efficient one. Since these bodies will be searched in known planetary systems, the advantage is that we can constrain the locations where to look for. In this regard, the technological stoppers at present are: the photometric precision that we can achieve and the time span of the monitoring. Kepler was sensitive enough to see Earth-sized moons in most cases (Kipping et al., 2017) and Galilean-sized moons when planets were stacked together into similar-type bins (Teachey et al., 2018). But detecting a single Europa-sized moon around a typical gas giant remains far too challenging for either Kepler or TESS. The photometric precision needed to detect such small bodies goes below the 1 part per million (ppm), around one order of magnitude smaller than Kepler, TESS, CHEOPS, and PLATO (a Moon-size body eclipsing a solar-like star induces 6 ppm). Reaching such precision we can measure precisely planet radii and consequently planet densities that will better constrain planet composition in the rocky regime.

Regarding the RV technique, the effect of exomoons on the RV of the star is several orders of magnitude smaller than the current foreseen precision of state-of-the-art instrumentation (like ESPRESSO). As proposed by Cabrera et al. (2007), the only foreseeable option to detect the radial velocity effect of an exomoon would be to see the planet-moon system as a binary (e.g., the Saturn-Titan system would have 2 m/s variations). This would require detecting the planet reflection with very high signal-to-noise ratio and fed its light to high-resolution spectrographs. The instrumentation necessary to reach these goals is now getting into place by combining instruments like SPHERE and ESPRESSO or GPI and HIRES but still few targets will be accessible even with the extremely large (40m-class) telescopes. In the case of trojans and comets, however, high-resolution spectroscopy can detect very small bodies (e.g., Leleu et al., 2017; Lillo-Box et al., 2018). The limitation here is the amount of available time. To detect a significant amount of these objects, large surveys or dedicated missions should be planned. In particular, multi-object high-resolution spectroscopy from space to achieve high-precision RVs in all-sky surveys (the radial velocity counterpart of the Kepler mission) would be a major step and would open a new era in the exploration of planetary systems.

## Summary

Minor bodies of our Solar System are thought to be crucial to understand life emergence and evolution on Earth, and they are a key components of our planetary system. Consequently their search around other stars (now that we have crossed the crystal roof of rocky planet detection) becomes a natural next step in exoplanet exploration. It is important to understand that despite having make an enormous improve in the study of planetary systems, the questions answered so far are still general and the starting point to many detailed studies, while being fundamental to the understanding of the emergence of life on Earth. As an example, the existence of the Moon is listed as one of the key factors for the stabilization of the precession over long time-scales and the evolution of life. Of course, the forthcoming missions planned to be launched (in particular JWST and TESS) in the following years will also push the limits of our understanding of planetary systems in all these questions. The next generation of space-based missions and technology should address the different limitations we will still have after the planned missions, such us better photometric precision (down to 1 ppm), multi-target high-precision radial velocity measurements or long-baseline optical and radio interferometry from the space.  These three requirements would be key to be able to continue our search for life in the Solar neighborhood and beyond.